\documentstyle[11pt,epsf]{article}
\begin{document}
\centerline{\bf ROTATIONAL BANDS AT THE LIMIT OF ANGULAR MOMENTUM}
\vspace{0.2cm}
\centerline{A. V. Afanasjev$^{1,2}$ and I. Ragnarsson$^1$}
\vspace{0.2cm}
\centerline{$^1$Department of Mathematical Physics, Lund 
Institute of Technology}
\centerline{PO Box 118, S-22100, Lund, Sweden}
\centerline{$^2$ Nuclear Research Center, Latvian Academy of Sciences}
\centerline{LV-2169, Salaspils, Miera str. 31, Latvia}
\vspace{0.5cm}
\setlength{\baselineskip}{0.4cm}

\begin{minipage}[h]{14.8cm}
\setlength{\baselineskip}{0.2cm}
{\bf Abstract.} {\it The properties of rotational bands 
at the limit of angular momentum are discussed on 
the example of {\rm smooth terminating bands} observed 
in the $A\sim 110$ mass region. The effective alignment 
approach is used for the study of their relative 
properties which provides additional insight into 
the properties of such bands.} 
\end{minipage}
\vspace{0.5cm}

With the new arrays of $\gamma$-detectors 
it has become possible to investigate rapidly rotating nuclei at the 
limit of angular momentum. The limit of angular momentum can 
be defined in terms of either the maximum spin at which a
nucleus exists as a bound system or the limited angular 
momentum within specific configurations (rotational bands). In 
the latter case we are dealing with the concept of {\it 
terminating bands}. These bands, which are collective 
at low spin, gradually lose their collectivity and 
exhaust their angular momentum content approaching a
pure single-particle ({\it terminating}) state of 
maximum spin. The existence of a maximum spin for a 
specific configuration is a manifestation of the 
finiteness of the nuclear many-fermion system where 
the angular momentum is built mainly from the contributions
of valence particles and valence holes, while the contribution
from closed shells is rather small or even negligible \cite{RXBR.86,A110}.
A typical situation is that the state of maximum spin has
oblate $(\gamma=60^{\circ})$ or prolate $(\gamma=-120^{\circ})$ 
non-collective shape. Keeping in mind that the termination 
of a rotational band takes place at  high spin where the pairing
correlations are of small importance, its configuration can be 
defined by the number of particles (holes) in different $j$-shells.
Since nucleons are fermions which obey the Pauli principle,
one valence particle in a $j$-shell contributes with $j\hbar$,
the next with $(j-1)\hbar$ etc.

However, in most cases this gradual interplay between 
collective and single-particle degrees of freedom is 
difficult to study in experiment. One typical situation, 
existing for example in the $A\sim 158$ mass region 
(see sect. 3 of \cite{A110} for an overview of experimental 
situation), is that although
the terminating states are seen in experiment, the rotational 
bands from which they originate go away from the yrast line
with decreasing spin and are thus difficult to observe over 
a large spin range. Another even more common situation is
that the rotational bands, which are yrast in some spin range, 
end up in terminating states residing well above the yrast
line. Only recent findings of {\it smooth terminating bands} 
in the $A\sim 110$ mass region \cite{Sb109,Rag95} observed 
over a wide spin range opened a real possibility for a detailed 
investigation of the terminating band phenomenon. Considering the spin range 
over which these bands are observed and the fact that in many cases 
their terminating states have been definitely or tentatively 
seen in experiment, this mass region indeed represents a 
unique laboratory for theoretical study of gradual interplay 
between collective and single-particle degrees of freedom 
within the nuclear system.

    A detailed theoretical investigation of smooth terminating
bands has been performed within the configuration-dependent 
shell-correction approach with the cranked Nilsson 
potential \cite{A110,Rag95}. The main ingredients of 
the approach used are:

\begin{itemize}

\item
Within the $N$-shells, virtual crossings between the 
single-particle orbitals are removed in an approximate
way and, as a result, smooth diabatic 
orbitals are obtained. An additional feature is that 
the high-$j$ orbitals in each $N$-shell are identified after
the diagonalization. As a result, it is possible to 
trace fixed configurations as a function of spin.

\item
The calculations are carried out in a mesh in the 
deformation space, $(\varepsilon_2,\varepsilon_4,\gamma)$. 
Then for each fixed configuration and each spin separately, 
the total energy of a nucleus is determined by a minimization 
in the shape degrees of freedom.

\item
Pairing correlations are neglected so the calculations
can be considered fully realistic for spins above 
$I \sim 20\hbar$ in the $A \sim 110$ mass region.  
An additional source of possible discrepancies between
experiment and calculations at low spin could be 
the one-dimensional cranking approximation used in 
the present approach.

\end{itemize}
Since most of configurations of interest in the 
$A\sim 110$ mass region are obtained by occupation 
of the $h_{11/2}$, $g_{7/2}$ and $d_{5/2}$ 
orbitals and by emptying of the proton $g_{9/2}$
orbitals, the shorthand notation $[p_1p_2,n]^{\alpha_{tot}}$
\cite{A110} is used for configuration labelling.
In this notation $p_1$ is the number of $g_{9/2}$ proton 
holes, $p_2$ the number of $h_{11/2}$ protons, $n$ the 
number of $h_{11/2}$ neutrons and $\alpha_{tot}$ is the 
total signature of the configuration.

\input epsf
\vspace*{-6.0cm}
\epsfxsize=0pt
\epsfbox{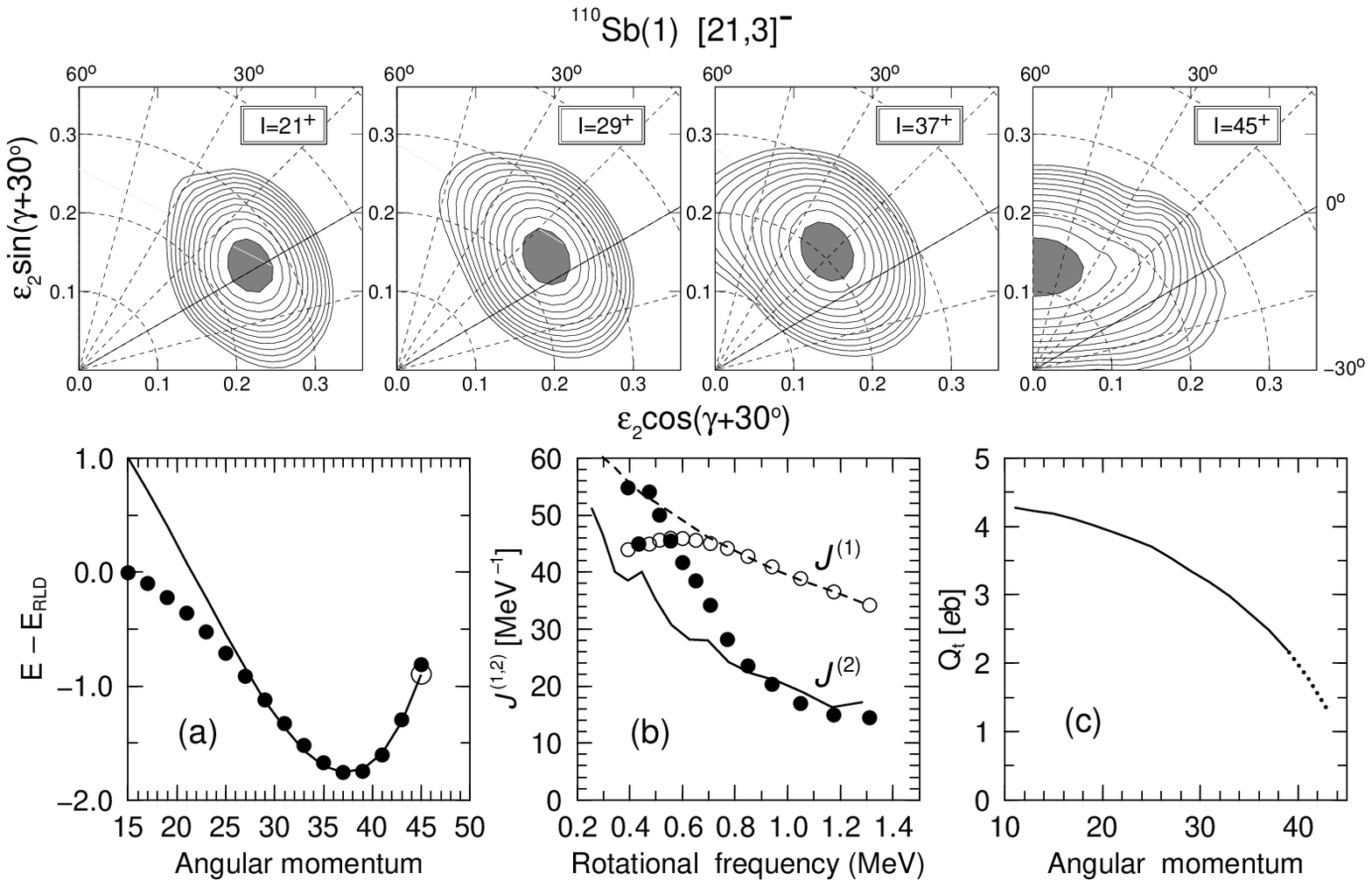}
\vspace*{-6.0cm}
Fig. 1. {\it Top row: Evolution of potential energy surfaces (PES) 
for fixed configuration $[21,3]^-$ assigned to band 1 
observed in $^{110}$Sb \protect\cite{Sb110}. PES are given in steps of $8\hbar$  
from near-prolate state at $I=21^+$ up to terminating state
at $I=45^+$. The minima are shaded. The energy difference
between two neighbouring equipotential lines is equal to
0.25 MeV and the last equipotential line corresponds
to 3.0 MeV excitation with respect of minimum. 
Bottom row: Comparison between theoretical results 
(curves without symbols) and experimental data (unlinked 
symbols) for band 1 in $^{110}$Sb. (a) Excitation energies 
given with respect to rigid rotor reference. 
Open circle indicates the predicted terminating state. (b) 
Kinematic $J^{(1)}$ and dynamic $J^{(2)}$ moments of inertia.
(c) The transition quadrupole moments $Q_t$ calculated
according to \protect\cite{A130}. For more detailed 
discussion of description of $Q_t$ in smooth terminating
bands see sect. 5.3.1 in \protect\cite{A110}.}
\vspace{0.5cm}

Essential physics related to smooth terminating bands can
be illustrated on the example of band 1 recently observed
in $^{110}$Sb \cite{Sb110}. This band was predicted
as an especially favoured intruder band in $^{110}$Sb, yrast 
over the spin range $25\hbar-45\hbar$, see Fig.\ 13 of \cite{A110}, 
one year before experimental data on this nucleus became
available. For spin values $I\geq 25\hbar$, where the role 
of pairing correlations is negligible,
the evolution of excitation energies as a function of 
spin within the band is in excellent agreement with 
theoretical predictions, see Fig.\ 1a. At spin $I\geq 38\hbar$, 
the $(E-E_{RLD})$ curve shows an upbend. This feature 
indicates that the last spin units before termination
are built at high energy cost; unfavoured termination. 
This high energy cost is determined mainly by \cite{A110}: 
(i) the difficulty to align 
the proton $g_{9/2}$ holes when they are surrounded
by aligned particles and (ii) the fact that the 
neutron $g_{7/2}\,d_{5/2}$ subshells are essentially
half-filled, i.e. the spin contribution from the last 
particles in these subshells is close to zero.
Experimentally, it is seen from the increase of the $\gamma$-ray 
energy spacings with increasing spin which implies that 
the dynamic  moment of inertia $J^{(2)}$ decreases to
values which are only a fraction of the rigid-body value. 
At rotational frequencies
$\hbar\omega>0.7$ MeV, where the role of pairing correlations
is negligible, the calculations reproduce this drop of 
$J^{(2)}$ rather well, see Fig.\ 1b. Furthermore, at these 
frequencies the calculated kinematic moment of inertia 
$J^{(1)}$ is in good agreement with experiment.
The smooth terminating bands show a continuous 
transition from high collectivity to a pure 
particle-hole (terminating) state which is 
illustrated in top row of Fig.\ 1. This transition 
is associatated with a gradual loss of collectivity 
as shown in Fig.\ 1c. Unfortunately, precise measurements 
of transition quadrupole moments $Q_t$ in smooth 
terminating bands are not available at present.

\vspace*{-11.7cm}
        \epsfxsize 14.0cm  \epsfbox{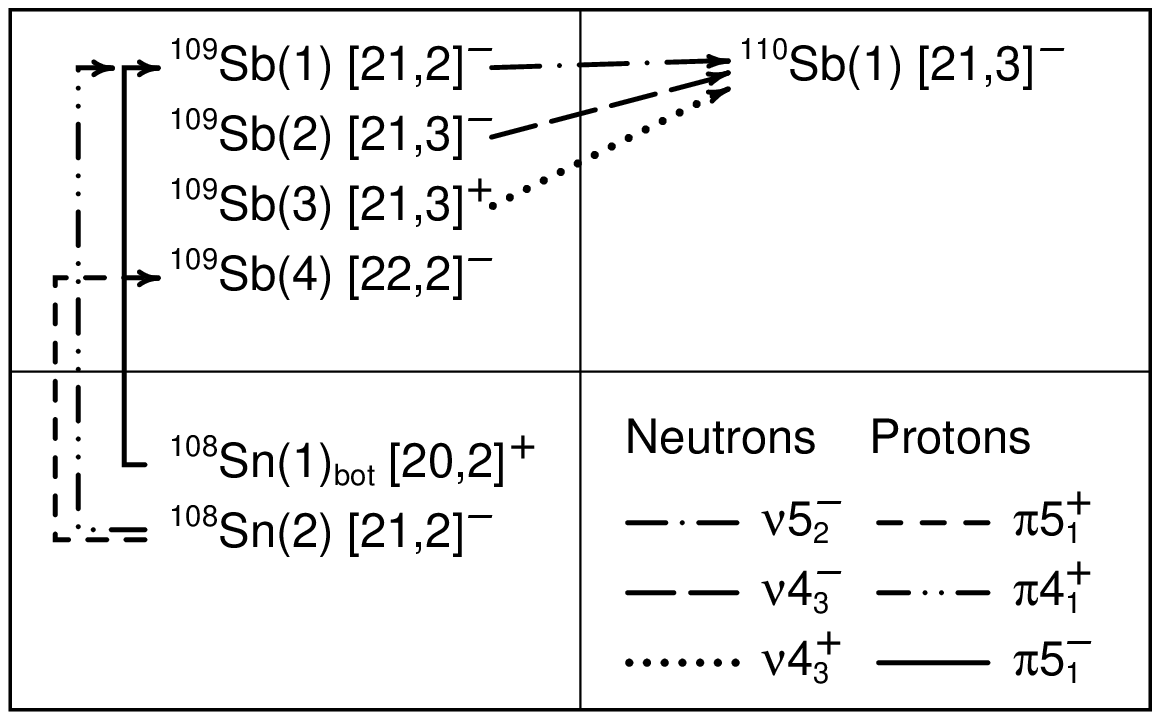}
\vspace*{-1.1cm}
Fig. 2. {\it Smooth terminating bands observed in $^{109,110}$Sb
and in $^{108}$Sn \protect\cite{Sb110,Sb109new,Sn108}. The assigned 
configurations are indicated after band
labels. The difference in the configurations of 
observed bands related to specific orbitals is indicated
by arrows of different types. The correspondence between
the type of arrow and orbital is given in right bottom 
panel. The orbitals are labeled by the main quantum
number, $N_{rot}$, by sign of signature $\alpha$ given 
as superscript and by the position of the orbitals 
within the specific signature group of the $N_{rot}$ shell 
given as subscript, counted from the $Z=50$ and $N=50$ 
spherical shell gaps. In this mass region, $N_{rot}=4$ 
corresponds to the $g_{7/2}\,d_{5/2}$ subshell and 
$N_{rot}=5$ corresponds to the $h_{11/2}$ subshell.}
\vspace{0.5cm}

One should note that in many cases, as for example 
in $^{109}$Sb, the observed bands are not linked to the 
low spin level scheme and the interpretation that they
are observed up to termination is partly based 
on comparison with the $(E-E_{RLD})$ curves 
for yrast and near-yrast configurations obtained in model
calculations. In other cases when such bands have been 
linked to the level scheme, as for example $^{108}$Sn(2) and 
$^{110}$Sb(1) bands, the transitions depopulating the 
terminating states have 
been established only tentatively \cite{Sb110,Sn108}.
As will be shown below, in such a situation the relative 
properties of unlinked and linked bands can be very useful
in order to establish if the present interpretation 
is consistent or not. We use an effective
alignment approach \cite{Rag91} succesfully applied for
an interpretation of superdeformed bands observed in
the $A\sim 150$ mass region employing present model
\cite{Rag91,Gd} and cranked relativistic mean field
theory \cite{ALR.97}. Effective alignment of two 
bands (A and B) is simply the difference between their spins 
at constant rotational frequency $\hbar\omega$ (or $\gamma$-transition
energy $E_{\gamma}$):
\begin{equation}
i^{B,A}_{eff}(\hbar\omega)=I^B(\hbar\omega)-I^A(\hbar\omega).
\end{equation}
Experimentally, $i_{eff}$ includes
both the alignment of the single-particle orbital and 
the effects associated with changes in deformation, pairing
etc. between two bands. This approach exploits the fact
that spin is quantized, integer for even nuclei and 
half-integer for odd nuclei and furthermore constrained
by signature. One should note that with the 
configurations and specifically the signatures 
fixed, the relative spins of observed bands
can only change in steps of $\pm 2\hbar \cdot n$ 
($n$ is integer number). 

\vspace*{-8.0cm}        
\leavevmode
        \epsfxsize 16.0cm  \epsfbox{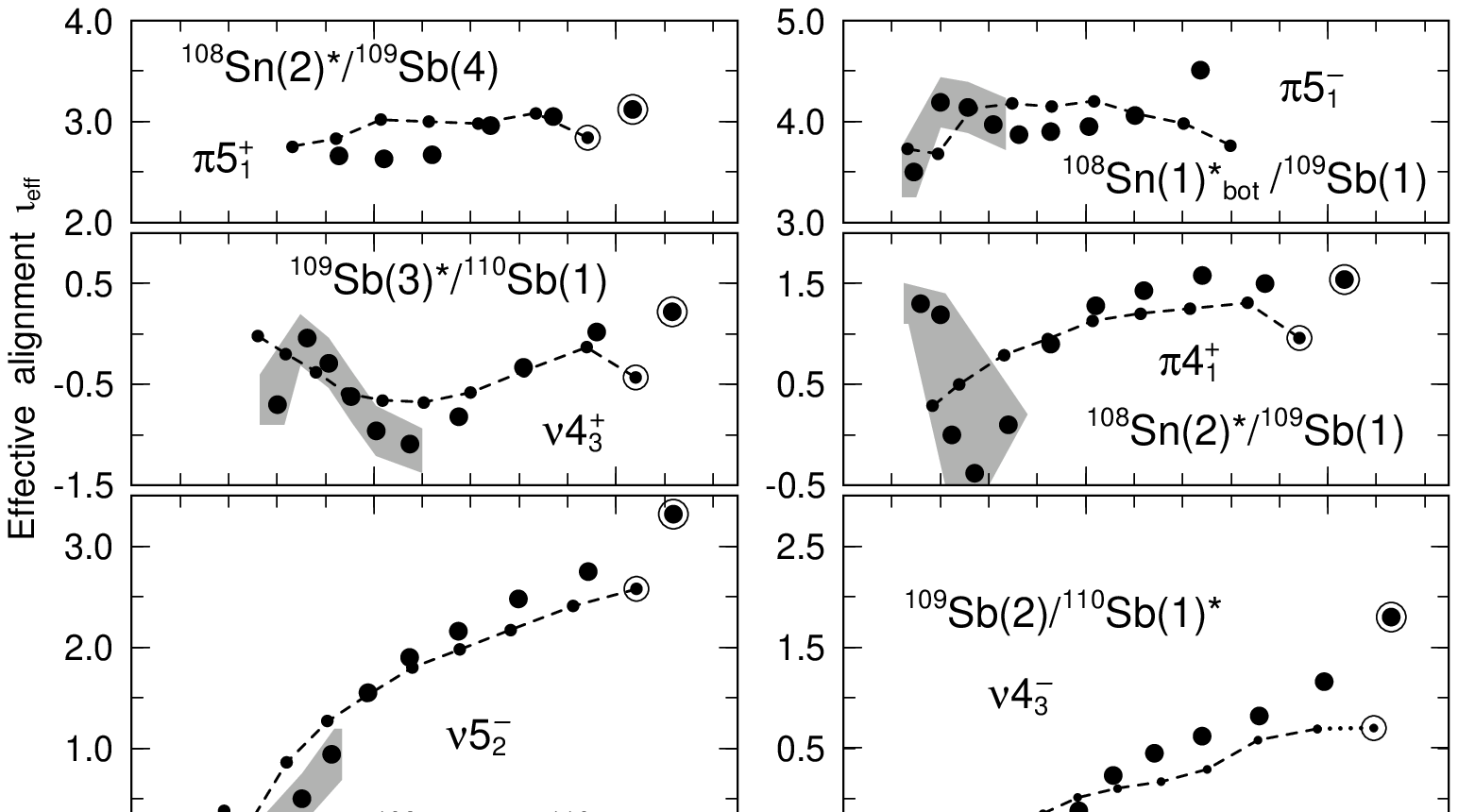}
\vspace*{5.0cm}
Fig. 3. {\it Effective alignments, $i_{eff}$ (in units
$\hbar$), extracted from experiment (unlinked large symbols) 
and corresponding calculated configurations (linked small 
symbols). The experimental effective alignment between 
bands A and B is indicated as ``A/B'' and it measures
the effect of additional particle.  
The $i_{eff}$ values are shown at the transition energies 
of the shorter band indicated by an asterisk {\rm ($^*$)}. The compared 
configurations differ by the occupation of the orbitals 
indicated on the panels. The points corresponding to 
transitions depopulating terminating states are 
encircled. The experimental points which, as follows from
the analysis of $J^{(2)}$ of compared bands, appears to be 
affected either by pairing interactions or by unpaired band 
crossings are shown on shaded background.}
\vspace{0.5cm}

In the present study, we exploit the fact that bands 
1 and 2 in $^{108}$Sn and band 1 in $^{110}$Sb are 
linked to the level scheme. The configuration assignment 
for observed bands, see Fig.\ 2, and the spin assignment 
for unlinked bands of $^{109}$Sb are the same as the 
one given in original articles 
\cite{Sb110,Sb109new,Sn108}. 
Considering the reasonable agreement between experiment and
calculations in the spin range where the pairing correlations
are of small importance, see Fig.\ 3, 
the following conclusions can be drawn:

\begin{itemize}

\item
The interpretation of smooth terminating bands 
observed in $^{109,110}$Sb and $^{108}$Sn 
\cite{Sb110,Sb109new,Sn108} based
on the features of experimental and theoretical
$(E-E_{RLD})$ curves is consistent with the present
study. The theoretical interpretation using the 
parabola-like behaviour of the $(E-E_{RLD})$ curves
of the bands which terminate in unfavoured way
appears reliable. In addition, the present analysis 
gives strong confidence in the spin assignment for 
unlinked bands 1-4 in $^{109}$Sb thus indicating that 
they have indeed been observed up to their terminating 
states.

\item
The effective alignment approach can be used 
also for the analysis of smooth terminating bands 
giving an additional and very sensitive 
tool for the interpretation of the observed bands.
However, compared with the case of superdeformed 
bands the changes in deformation between 
two bands play a much more important role and, 
as a consequence, the  effective alignment does not 
necessary reflect the alignment of single-particle 
orbital. This can be illustrated by the fact that 
the effective alignment in the 
$^{108}$Sn(2)$_{bot}^*$/$^{109}$Sb(1) pair
is $\approx 1\hbar$ lower than pure single-particle 
alignment of the lowest $\pi h_{11/2}$ orbital, see
Fig.\ 10 in \cite{A110}. 

\item
It is more difficult to reproduce the $i_{eff}$ 
values for the transitions depopulating 
terminating states. These transitions link 
the states having largest difference in equilibrium 
deformation between two neighbouring states 
within a band. For example, in this mass 
region this difference could reach 
$\sim 30^{\circ}$ for $\gamma$-deformation.
As a result, these $i_{eff}$ values are more
sensitive both to the accuracy of reproduction 
of prolate-oblate energy difference in the
liquid-drop part and the parametrization of the
Nilsson potential than the $i_{eff}$ values for 
transitions at lower spin.

\end{itemize}

Detailed theoretical investigations within the configuration-dependent
shell-correction approach performed recently \cite{ARprep,A80a} indicates 
that similar bands could be observed in the $A\sim 70-80$ mass region. 
One interesting feature in this region is the way in which some of 
these bands can evolve at $I \geq I_{max}$. At low spins in the cranked 
harmonic oscillator, configurations
can be defined by the number of particles in the different $N$-shells and
the maximum spin within each $N$-shell is easily deduced. At large deformations, 
$\beta \sin(\frac{\pi}{3}-\gamma)\geq 0.511$, however, the $N$-shells mix strongly 
and then with increasing rotational frequency, more angular momentum is
continuously pumped into the configuration so that it can continue as a collective 
rotation beyond $I_{max}$ \cite{Tr}. In a realistic nuclear potential for
the nuclei in the $A \sim 70-80$ mass region, the $f_{7/2}$ high-$j$ subshell
splits off from the other $N=3$ subshells and, at low spin, the number
of particles in $f_{7/2}$ and in other $N=3$ subshells can be specified
separately. With increasing angular momentum, however, all the $N=3$ subshells
mix very strongly and according to our calculations, such a configuration
can, in a similar way as for the pure oscillator, continue as collective beyond 
the maximum spin $I_{max}$ as defined from the distribution of the
valence nucleons over the $j$-shells
at low spin. One should note that in a realistic nuclear potential, such a 
situation arises at considerably smaller deformation compared with
the strong mixing of $N$-shells in the pure oscillator. For example, 
at low spin, the configurations of interest in the $A\sim 70-80$
mass region are triaxial close to oblate with $\gamma=-40^{\circ}$ 
and $\varepsilon_2\approx 0.3$.
\vspace{0.5cm}

{\it We are grateful for financial support from the 
Royal Swedish Academy of Sciences, from the Crafoord 
Foundation (Lund, Sweden) and from the Swedish Natural
Science Research Council.}

\end{document}